 \documentclass[times,review,preprint,authoryear]{elsarticle}

\RequirePackage{geometry}
\usepackage{framed,multirow}

\usepackage{amsmath}
\usepackage{amssymb}
\usepackage{latexsym}
\usepackage{natbib}
\usepackage{hyperref}
\usepackage{graphicx}
\usepackage{epstopdf}
\usepackage{psfrag}
\usepackage{overpic}
\usepackage{CJK}
\usepackage{overpic}
\usepackage{tikz}
\usepackage{bm}
\usepackage{subfigure}
\usepackage{url}
\usepackage{xcolor}
\usepackage{diagbox}
\usepackage{multirow}
\definecolor{newcolor}{rgb}{.8,.349,.1}

\newcommand\Rey{\mbox{\textit{Re}}}  % Reynolds number
\newcommand{\diff}{\mathrm{d}}

\begin{document}

%\verso{Pirozzoli \& Orlandi}

\begin{frontmatter}

\title{Natural grid stretching for DNS of wall-bounded flows}

\author[]{Sergio Pirozzoli\corref{cor1}}

\cortext[cor1]{Corresponding author:
  Tel.: (+39) 06 44585202;
  fax: (+39) 06 44585250;}
\ead{sergio.pirozzoli@uniroma1.it}

\author[]{Paolo Orlandi}

\address{Dipartimento di Ingegneria Meccanica e Aerospaziale, Sapienza Universit\`a di Roma, Via Eudossiana 18, Roma 00184 , Italy}
%\received{1 May 2013}
%\finalform{10 May 2013}
%\accepted{13 May 2013}
%\availableonline{15 May 2013}
%\communicated{S. Sarkar}

\begin{abstract}
We propose a natural stretching function for DNS of wall-bounded flows, which blends
uniform near-wall spacing with uniform resolution in terms of Kolmogorov units
in the outer wall layer. Numerical simulations of pipe flow are used to educe optimal value
of the blending parameter and of the wall grid spacing which guarantee accuracy and computational efficiency
as a results of maximization of the allowed time step. Conclusions are supported by DNS carried 
out at sufficiently high Reynolds number that a near logarithmic layer is the mean velocity profile 
is present. Given a target Reynolds number, we provide a definite prescription 
for the number of grid points and grid clustering needed to achieve 
accurate results with optimal exploitation of resources.
\end{abstract}

%\begin{keyword}
%\KWD Direct numerical simulation \sep Wall turbulence 
%\end{keyword}

\end{frontmatter}

%\linenumbers

%% main text

\section{Introduction}

Direct numerical simulation (DNS) of wall-bounded flows is by now an established practice,
started from the pioneering work of \citet{kim_87} for channel flow.
Although meshing is not a challenging issue given the simple topology 
of canonical flows to which DNS is currently limited, 
the mesh parameters significantly affect computational accuracy and efficiency.
It is generally acknowledged~\citep[e.g.][]{lee_15} that mesh spacings in the order of 
ten wall units in the streamwise direction and five in the spanwise direction 
are sufficient in pseudo-spectral calculations to achieve good resolution 
of the buffer-layer energy-containing eddies, namely streaks 
and associated quasi-streamwise vortices.
The buffer layer is especially important as the topology of eddies changes from
sheet-like near the wall to rod-like away from it, corresponding to the 
inflectional point of the wall-normal velocity variance profile~\citep{orlandi_13}.
Finite-difference schemes require similar or slightly higher number of grid
points~\citep{bernardini_14}, to achieve the same quality of results.
More disputable is the selection of the mesh properties in the 
wall-normal direction, which is strongly anisotropic for the flow,
and for which no rule is consolidated yet.
In fact, different authors of state-of-art DNS have used vastly different
mapping functions, and the selection of the total number of grid points is
mainly a matter of personal experience and feeling.
Another important issue in the design of modern DNS is computational efficiency.
In fact, it turns out that the admissible time step is strongly affected by the
wall-normal distribution of the grid points, and changing the mapping
function can yield substantial saving of computer time, with little or no loss 
of accuracy. While some inefficiency is forgiven in small-scale DNS carried out
on local computer clusters, this is clearly not allowed in leading-edge numerical simulations
exploiting huge computational resources. 
The purpose of this paper is to provide the community of DNS of wall-bounded flows 
with a tailored mapping function and definite grid point number estimates, 
so as to satisfy natural resolution requirements and at the 
same time to provide maximum computational performance.
Although the forthcoming discussion is mainly focused on the case
of turbulent pipe flow, other canonical cases can also be handled with no or minimum
modifications, as plane channel and boundary-layer flows.

\section{Wall-normal stretching functions}
\label{sec:stretching}

Considerations about the multi-scale nature of wall-bounded turbulence lead to conclude
that several constraints shall be satisfied for effective design 
of the wall-normal mapping function:
i) the first off-wall grid node (say $\Delta y_w^+$) shall be placed close enough that
the severe velocity gradients occurring in that region are resolved, which 
requires $\Delta y_w^+ \lesssim 1$;
ii) grid points should be conveniently clustered within the buffer layer (say, $y^+ \le 50$) 
which is the most anisotropic region of the flow, and in which most intense phenomena occur;
iii) the spacing in the outer part of the wall layer, in which turbulence is not far 
from isotropic, should be proportionate to the local Kolmogorov length scale.
Synthetic information about previous DNS studies of channel and pipe flows is provided
in Table~\ref{tab:params}, where $\Rey_{\tau} = u_{\tau} \delta / \nu$ 
(with $u_{\tau}$ the friction velocity, $\delta$ either the channel half-height or the pipe radius
or the boundary layer thickness, and $\nu$ the fluid kinematic viscosity) is the friction Reynolds number,
$N_y$ is the number of collocation points in the wall-normal direction, and $N_{buf}$ is the
number of grid points within the buffer layer.
As can be seen, different studies rely on different mapping functions (see \citet{orlandi_00}
for an overview of classical ones), different near-wall resolutions, and even very different 
total number of points for similar $\Rey_{\tau}$.
In this respect it should be noted that standard spectral methods only
allow for cosine stretching in the vertical direction to exploit Chebyshev transform,
and alternate mappings can only be accommodated by changing the
numerical treatment of the wall-normal direction.
For instance, \citet{hoyas_06} used sixth-order compact differencing,
whereas \citet{lee_15} used a B-spline collocation method.
On the other hand, the finite-difference method allows use of arbitrary mappings. 

\begin{table}
\begin{center}
\begin{tabular}{cccrrcrc}
\hline	
Flow & Reference & Stretching function & $\Rey_{\tau}$ & $N_y$ & $\Delta y^+_w$ & $N_{buf}$ & Symbol \\
\hline
\multirow{13}{*}{Channel} & \citet{kim_87}                  & Cosine                &  180 &  64 & 0.05   & 31  & $\times$ \\
\cline{2-8}
& \multirow{3}{*}{\citet{hoyas_06}} & \multirow{3}{*}{NA} &  550 & 128 & 0.041  & 36 & \\
&                                   &                         &  934 & 192 & 0.031  & 40 & \\
&                                   &                         & 2004 & 317 & 0.32   & 31 & $\square$ \\
\cline{2-8}
& \multirow{4}{*}{\citet{lee_15}} & \multirow{4}{*}{Cosine/splines} &  182 &  96 & 0.074  & 48 & \\
&                                 &                                 &  544 & 192 & 0.019  & 56 & \\
&                                 &                                 & 1000 & 256 & 0.019  & 55 & \\
&                                 &                                 & 5186 & 768 & 0.50   & 54 & \huge{$\circ$} \\
\cline{2-8}
& \citet{lozano_14}               & NA                    & 4180 & 540 & 0.32  & 31 & \\
\cline{2-8}
& \multirow{4}{*}{\citet{pirozzoli_16}} & \multirow{4}{*}{Error function} &  548 &  192 & 0.06 & 79 & \\
&                               &                       &  995 & 256 & 0.018  & 79 & \\
&                               &                       & 2017 & 384 & 0.26  & 83 & \\
&                               &                       & 4088 & 512 & 0.38  & 73 & \large{$\triangle$} \\
\hline
\multirow{10}{*}{Pipe} & \multirow{2}{*}{\citet{wu_08}}  & \multirow{2}{*}{NA}    &   182 & 256 & 0.17 & 150 & \\
&                               &                       &  1142 & 300 & 0.41 & 69 & $\triangledown$ \\
\cline{2-8}
& \multirow{4}{*}{\citet{chin_14}}& \multirow{4}{*}{NA}    &   180 &  80 & 0.50 & NA & \\
&                               &                       &   500 & 160 & 0.07 & NA & \\
&                               &                       &  1002 & 192 & 0.6  & NA & \\
&                               &                       &  2003 & 320 & 0.35 & NA & \\
\cline{2-8}
& \multirow{4}{*}{\citet{ahn_15}}& \multirow{4}{*}{NA}     &   180 & 257 & 0.165 & NA & \\
&                               &                       &   544 & 279 & 0.176 & NA & \\
&                               &                       &   934 & 301 & 0.33  & NA & \\
&                               &                       &  3008 & 901 & 0.36  & NA & \\
\hline 
Boundary layer & \citet{schlatter_10b}          & NA              & 1271 & 212 & 0.033  & 40 & \huge{$\diamond$} \\
\hline 
\end{tabular} 
\end{center}
\caption{List of mesh parameters for reference DNS studies. $\Delta y_w$ is the distance of the first off-wall node, and 
$N_{buf}$ is the number of grid points within $y^+ \le 50$.}
\label{tab:params}
\end{table}

We believe that, given the near universality of wall-bounded flows,
a universal treatment of the stretching function is possible and appropriate.
We then reason as follows. First, following considerations of \citet{hoyas_06},
we believe that the grid spacing in the outer layer should be selected to be
proportional to the local Kolmogorov length scale, say $\eta$. 
By definition $\eta^+ = {\varepsilon^+}^{-1/4}$, with $\varepsilon$ the local
dissipation rate, and $+$ denoting wall units. Since under 
equilibrium conditions in the log layer $\epsilon^+ \sim 1/y^+$, 
it follows~\citep{jimenez_18} that
\begin{equation}
\eta^+ \approx c_{\eta} {y^+}^{1/4} , \label{eq:etap}
\end{equation}
with $c_{\eta} \approx 0.8$, which is consistent with all available DNS of channel and pipe flow,
as we have directly checked.
Hence, the first requirement which we set is that the local
grid spacing in the outer layer should be $\Delta y^+ = \alpha \eta^+$, 
with $\alpha$ controlling adequate resolution of the dissipative eddies.
The choice $\alpha=1.5$ yields a resolution in spectral space $k_{max} \eta \approx 2$
(where $k_{max} = \pi / \Delta y$ is the maximum resolved wavenumber) which is regarded
to be sufficient in numerical simulations of isotropic turbulence~\citep{jimenez_98},
and similar to the resolution used in channel flow by \citet{hoyas_06}.
Thus, let $j$ be the wall-normal grid index (momentarily assumed to be 
continuous for convenience), we require that
\begin{equation}
\Delta y^+ = \frac{\diff y^+}{\diff j} = \alpha c_{\eta} {y^+}^{1/4}, \label{eq:dypdj}
\end{equation}
which upon integration yields
\begin{equation}
y^+(j) = \left( \frac 34 \alpha c_{\eta} j \right)^{4/3}, \label{eq:yplog}
\end{equation}
which defines the mesh stretching in the outer wall layer.
Next to the wall, in the viscous sublayer the mean velocity gradient is nearly constant up to
$y^+ \approx 5$, and use of uniform spacing is appropriate, hence
\begin{equation}
y^+(j) = \Delta y^+_w \cdot j. \label{eq:ypwall}
\end{equation}
Whereas experience from most previous DNS suggests that a reasonable value of the wall grid spacing
be $\Delta y^+_w \approx 0.1$, its influence on the DNS statistics will be herein discussed.
A smooth blending between the near-wall mapping \eqref{eq:yplog} and the outer-layer mapping \eqref{eq:yplog}
is further assumed, to yield
\begin{equation}
y^+(j) = \frac 1{1+(j/j_b)^2} \left[ \Delta y^+_w j + \left( \frac 34 \alpha c_{\eta} j \right)^{4/3} (j/j_b)^2 \right], \label{eq:mapping}
\end{equation}
where the parameter $j_b$ defines the grid index at which transition between the 
near-wall and the outer mesh stretching should take place, whose choice will also be discussed in detail.
Straightforward differentiation of Eqn.~\eqref{eq:mapping} also yields the local grid spacing,
\begin{equation}
\Delta y^+ = \frac{\diff y^+}{\diff j} = \frac 1{\left( 1+(j/j_b)^2 \right)^2} \left[ \left( 1-(j/j_b)^2 \right) \Delta y^+_w + \frac 23 \left( \frac 34 \alpha c_{\eta} \right)^{4/3} \frac {j^{7/3}}{j_b^2} \left( 5 + 2 (j/j_b)^2\right) \right] . \label{eq:dyp}
\end{equation}
Evaluating Eqn.~\eqref{eq:mapping} at the edge of the wall layer yields the
number of grid points along the vertical direction as is an implicit function of 
$\Rey_{\tau}$ (as it should be), and of the stretching parameters $j_b$, $\Delta y^+_w$, $\alpha$.
However, under the assumption $j>>1$, hence at sufficiently high Reynolds number
that the number of points within the buffer layer becomes small
compared to those in the outer layer, Eqn.~\eqref{eq:mapping} yields 
\begin{equation} 
\Rey_{\tau} = \left( \frac 34 \alpha c_{\eta} N_y \right)^{4/3}, \label{eq:retau}
\end{equation} 
whence the number of necessary grid points to achieve a given (large) $\Rey_{\tau}$ can be estimated,
\begin{equation}
N_y = \frac{4}{3 \alpha c_{\eta}} \Rey_{\tau}^{3/4}, \label{eq:Ny}
\end{equation}
where the nearest integer should be taken for practical purposes.
Representative numbers are given in Table~\ref{tab:Ny} .
\begin{table}
\begin{center}
\begin{tabular}{ccccccc}
\hline	
$\Rey_{\tau}$ & 500 & 1000 & 2000 & 5000 & 10000 & 20000 \\
$N_y$ & 140 & 237 & 399 & 793 & 1333 & 2242 \\
\hline
\end{tabular}
\end{center}
\caption{Estimated number of grid points in wall-normal direction for DNS of wall turbulence,
according to the asymptotic formula~\eqref{eq:Ny}}.
\label{tab:Ny}
\end{table}
Equation~\eqref{eq:Ny} is quite interesting as it suggests that the total number of grid points
for DNS of wall-bounded turbulence should scale as $N_{xyz} \sim \Rey_{\tau}^{11/4}$,
whereas common estimates suggest $N_{xyz} \sim \Rey_{\tau}^3$, on the grounds that the
thickness of the viscous wall region determines the smallest scales throughout the wall layer~\citep{reynolds_90}. 
Available channel and pipe flow data further suggest that for
$\Rey_{\delta} \gtrsim 10^5$, $\Rey_{\tau} \sim \Rey_{\delta}^{0.92\--0.94}$ (where $\Rey_{\delta} = 2 u_b \delta / \nu$
is the bulk Reynolds number), thus yielding an estimated total number of grid points,
$N_{xyz} \sim \Rey_{\delta}^{2.5\--2.6}$. 

%\begin{table}
%\begin{center}
%\begin{tabular}{cccc}
%\hline	
%Flow & Reference & $\Rey_{\tau}$ & Symbol \\
%\hline
%\multirow{4}{*}{Channel}  & \citet{kim_87}   &  180 & $\times$ \\ 
%                          & \citet{hoyas_06} & 2004 & $\square$ \\ 
%                          & \citet{lee_15}   & 5186 & \huge{$\circ$} \\ 
%                          & \citet{pirozzoli_16} & 4088 & \large{$\triangle$} \\ 
%\hline
%Pipe                      & \citet{wu_08}    & 1142 & $\triangledown$ \\
%Boundary layer            & \citet{schlatter_10b} & 1271 & \huge{$\diamond$} \\
%\hline 
%\end{tabular} 
%\end{center}
%\caption{List of DNS datasets under scrutiny for grid resolution analysis.}
%\label{tab:symbols}
%\end{table}

\begin{figure}
 \begin{center}
  (a) \includegraphics[width=7.0cm]{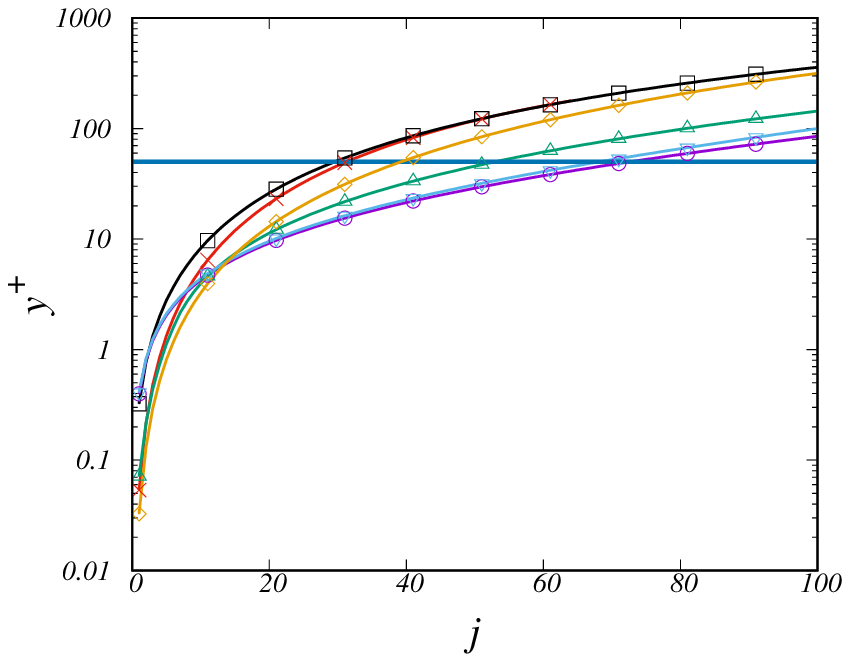}
  (b) \includegraphics[width=7.0cm]{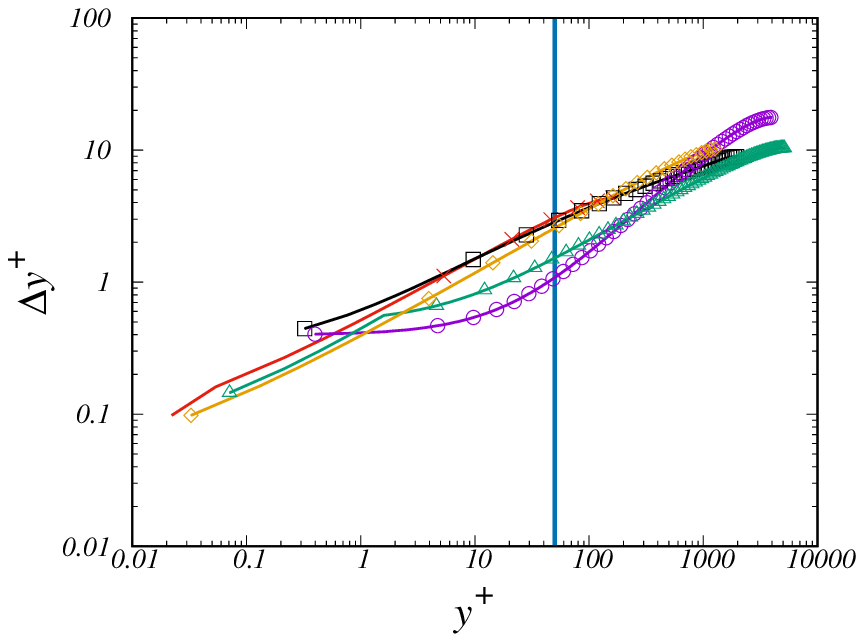} \\
  (c) \includegraphics[width=7.0cm]{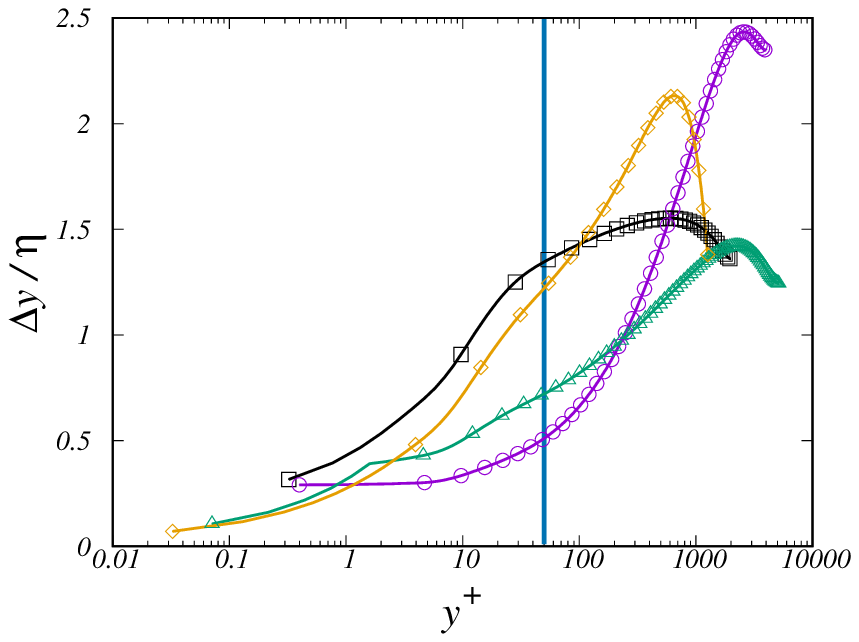}
\caption{Mapping functions (a) and corresponding grid spacing distributions in wall units (b) and in Kolmogorov units (c) for representative DNS of wall-bounded flows. Symbols as given in Table~\ref{tab:params}. Straight horizontal and vertical lines denote the edge of the buffer layer ($y^+=50$).}
  \label{fig:mappings}
 \end{center}
\end{figure}

The diversity of grid mappings used for DNS of wall-bounded flows is reflected in figure~\ref{fig:mappings} ,
showing the distribution of grid points for representative simulations. The mappings differ in several respects, 
including position of the first wall point, and number of points in the buffer layer, whose 
edge is marked with a horizontal line. Specifically, in the DNS of \citet{kim_87} and \citet{hoyas_06}
the grid points are more clustered towards the wall, at the expense 
of having limited number of points in the buffer layer, about thirty. Other DNS~\citep{wu_08,pirozzoli_16}
have larger wall spacing and more points in the buffer layer, up to seventy. Other cases~\citep{lee_15,schlatter_10b}
fall in between. This difference is appreciated in panel (b), showing the mesh spacing as a function 
of the wall distance. Whereas most DNS  have a spacing of 2-3 wall units at the edge of the buffer layer,
other have nearly uniform spacing in the viscous sublayer, and spacing of about one wall unit at the
edge of the buffer layer. To have a perception for the grid resolution in the outer layer,
in panel (c) we show the mesh spacing normalized by the local Kolmogorov length scale,
limited to those cases in which the latter is available.
The effective resolution of most DNS in the outer layer is about 1.25-1.5 Kolmogorov units,
and a bit poorer in the case of the DNS of \citet{schlatter_10b} and \citet{pirozzoli_16}.

\section{Numerical tests}

\subsection{Effect of the $j_b$ parameter}
\label{sec:jb}

\begin{figure}
 \begin{center}
  (a) \includegraphics[width=7.0cm]{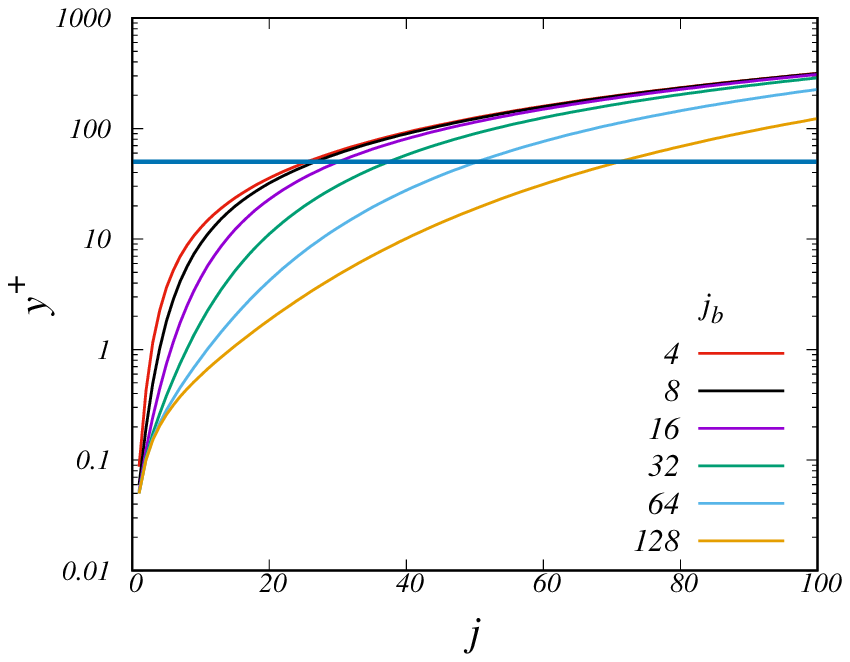}
  (b) \includegraphics[width=7.0cm]{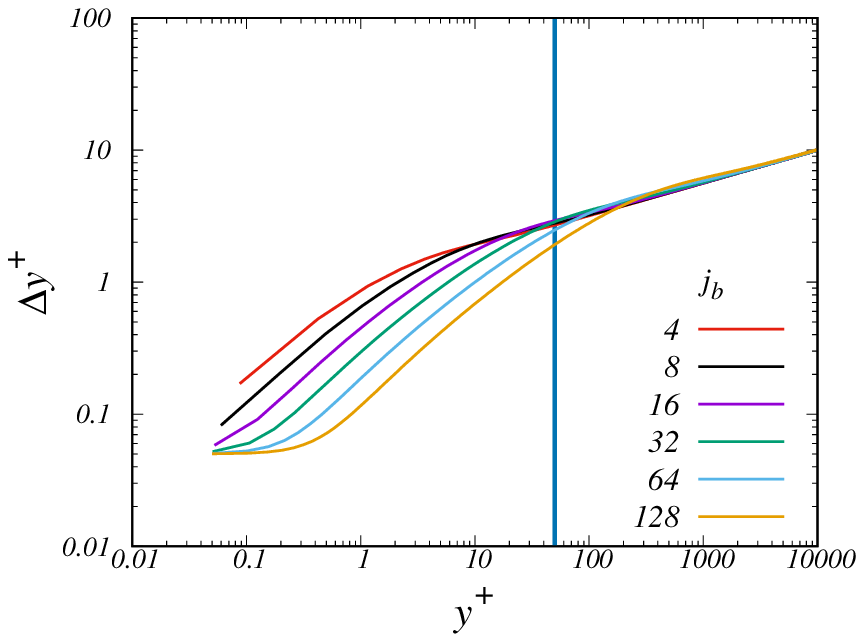}
% (a) \includegraphics[width=7.0cm     ]{FIGURES/yp_new_all.eps}
% (b) \includegraphics[width=7.0cm,clip]{FIGURES/dyp_new_all.eps}
\caption{Mapping functions (a) and corresponding grid spacing distributions (b), according to equations~\eqref{eq:mapping}, \eqref{eq:dyp}, as a 
function of $j_b$, assuming $\Delta y_w^+=0.05$, $\alpha=1.25$. Straight horizontal and vertical lines denote the edge of the buffer layer ($y^+=50$).} 
  \label{fig:mappings_new}
 \end{center}
\end{figure}

%\begin{table}
%\begin{center}
%\begin{tabular}{ccc}
%\hline	
%$J_b$ & $\Delta y^+_w$ & $C_f$ \\
%\hline
%4 & 0.05   &  9.34512541E-03 \\
%8 & 0.05   &  9.33915377E-03 \\
%16 & 0.05  &  9.38012078E-03 \\
%32 & 0.05  &  9.39425360E-03 \\
%64 & 0.05  &  9.43240337E-03 \\
%128 & 0.05 &  9.45049431E-03 \\
%16 & 0.01  &  9.38537755E-03 \\
%16 & 0.10  &  9.33295914E-03 \\
%16 & 0.50  &  9.31028838E-03 \\
%16 & 1.0   &  9.32362782E-03 \\
%16 & 2.0   &  9.44099001E-03 \\
%16 & 5.0   &  9.55460087E-03 \\
%\hline  
%\end{tabular}
%\end{center}
%\caption{Friction factor for several combination of the stretching parameters}
%\label{tab:f}
%\end{table}

The stretching functions herein designed along with the associated grid spacing distributions 
are shown for various $j_b$ in figure~\ref{fig:mappings_new},
where we assume $\Delta y^+_w=0.05$, $\alpha=1.25$.
As intended, the $j_b$ parameter controls the number of grid points within the buffer layer, changing
from about $25$ to about $70$ as $j_b$ ranges between 4 and 128. 
The near-wall spacing is increasing at low $j_b$, whereas
for all cases the grid spacing at the edge of the buffer layer is $\Delta y^+ \approx 3$, 
and $\Delta y^+ \approx 2$ for $j_b=128$.
Comparison with figure~\ref{fig:mappings} shows
that change of $j_b$ allows to basically cover the range of stretching functions used in previous studies.

\begin{figure}
 \begin{center}
  (a) \includegraphics[width=7.0cm,clip]{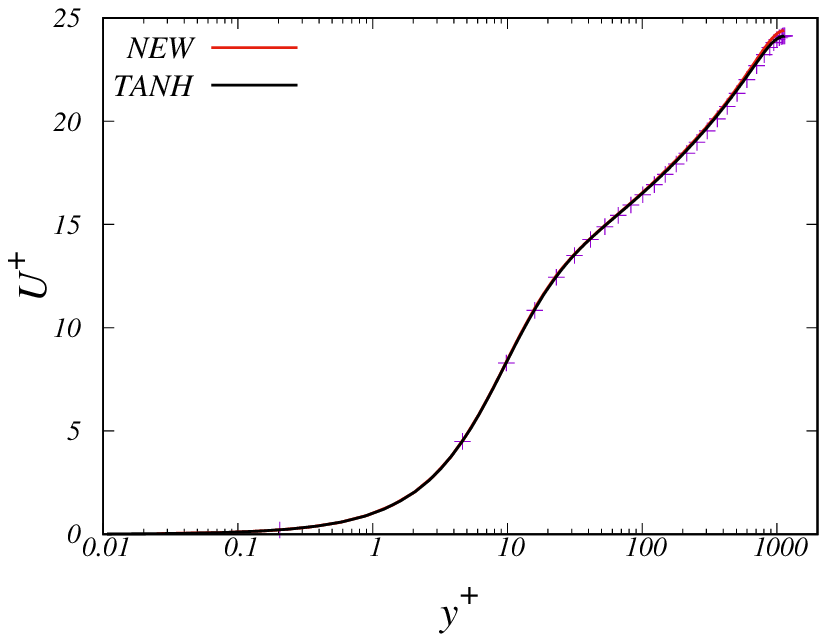}
  (b) \includegraphics[width=7.0cm,clip]{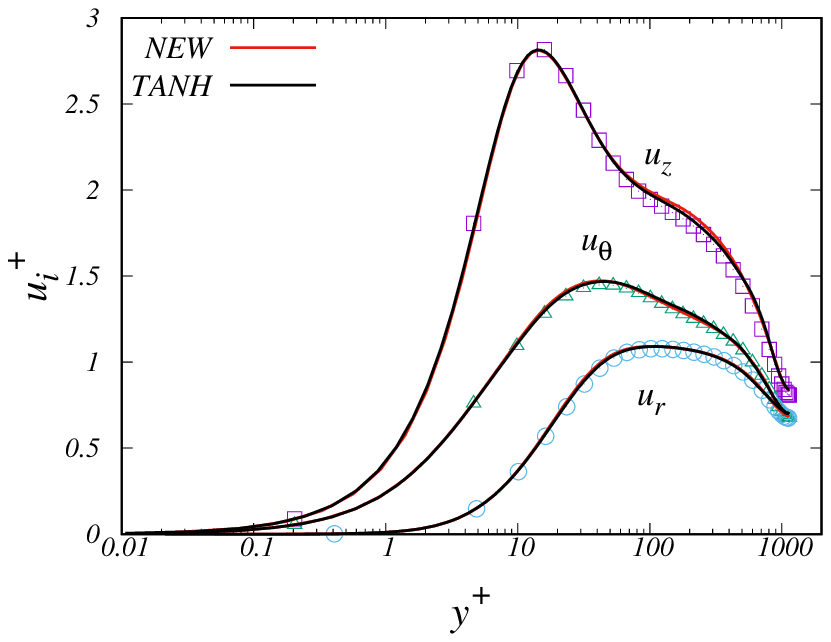} \\
  (c) \includegraphics[width=7.0cm,clip]{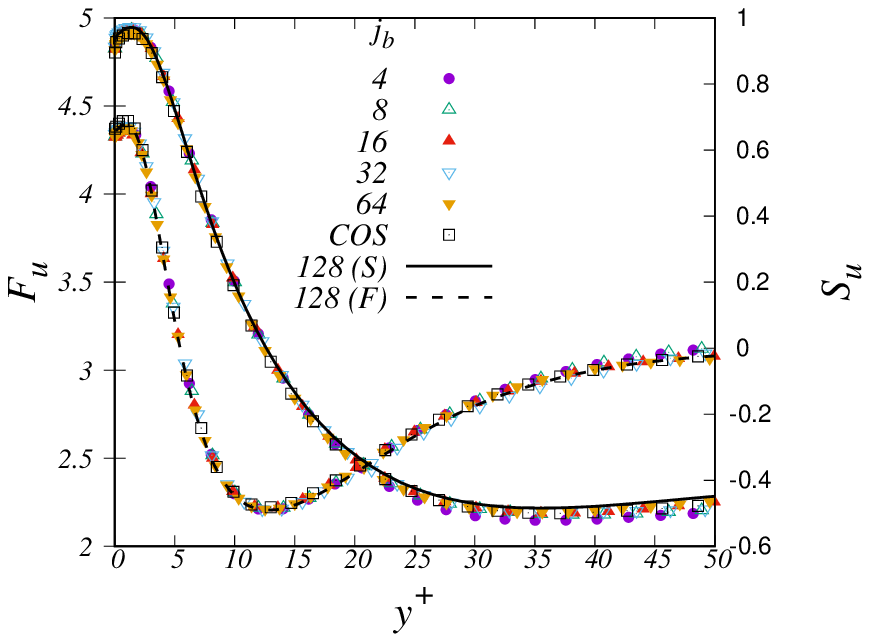}
% (d) \includegraphics[width=7.0cm,clip]{FIGURES/PIPE/dissip.eps}
  (d) \includegraphics[width=7.0cm,clip]{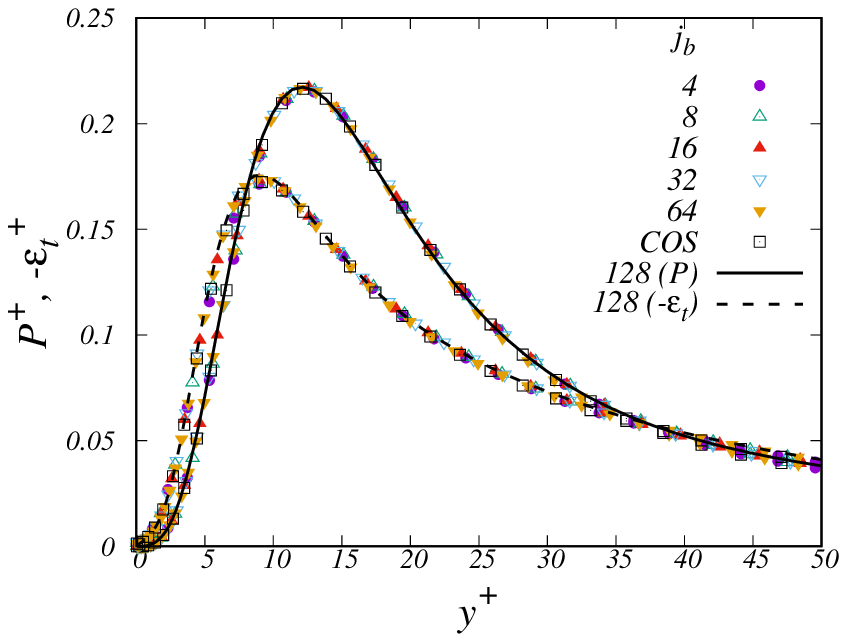}
% (d) \includegraphics[width=7.0cm,clip]{FIGURES/PIPE/dissip_all.eps}
\caption{DNS of pipe flow at $\Rey_{\delta}=5300$, effect of $j_b$ stretching parameter: 
profiles of (a) mean velocity, (b) r.m.s. velocity fluctuations (axial, $u_z$; wall-normal, $u_r$; azimuthal, $u_{\theta}$), 
(c) skewness and flatness of $u_z$,
and (d) turbulence kinetic energy production ($P=-\overline{u_r u_z} \, \diff U / \diff y$, solid lines),
and total dissipation ($\varepsilon_t = \nu \overline{u_i \nabla^2 u_i}$, dashed lines). 
Symbols in panels (a, b) denote velocity statistics of \citet{wu_08}. 
Lines in panels (c, d) denote numerical results obtained with $j_b=128$, which
are used as a reference. The square symbols denotes results obtained using the cosine stretching function, with $N_y=64$.}
  \label{fig:stats_Re5300}
 \end{center}
\end{figure}

In order to show how the choice of $j_b$ impacts the quality of numerical results and the 
involved computational effort we have carried out a series of numerical experiments of pipe flow at modest 
Reynolds number ($\Rey_{\delta}=5300$), using a well-established solver~\citep{verzicco_96,orlandi_97},
modified with implicit treatment of the azimuthal convective terms~\citep{akselvoll_96,stevens_13}.
The rationale is that, since the buffer-layer dynamics is weakly affected by Reynolds
number variations, the results of the analysis can be extrapolated to higher Reynolds number.
In all the simulations the computational domain is $15$ pipe radii long,
and $256$ grid points are used in the axial and azimuthal directions,
with corresponding grid spacings $\Delta x^+ \approx 10.7$, $R^+ \Delta \theta = 4.5$.
In these exploratory simulations the value of $N_y$ is varied from case to
case, according to equation~\eqref{eq:mapping}, from $N_y=66$ for $j_b=4$ to $N_y=116$ for $j_b=128$.
However, differences in the total number of points would become vanishingly small 
at higher $\Rey$, as reflected in the asymptotic formula \eqref{eq:Ny}. 
The resulting flow statistics are shown in figure~\ref{fig:stats_Re5300},
along with reference data of \citet{wu_08}, who reported a friction Reynolds number
$\Rey_{\tau}=181.4$. Here, we find that $\Rey_{\tau}$ ranges between $181.1$ and $182.1$ as $j_b$
varies, hence the impact on frictional drag is less than $0.5 \%$. 
The impact is also small on the main 
flow statistics, including inner-scaled mean velocity profiles (panel (a)) 
and velocity fluctuations intensities (panel (b)),
although limited scatter and slight differences with respect to the reference data
are visible in the axial turbulence intensity towards the pipe axis.
The higher-order moments of $u$ (panel(c)) show that lower resolution
implies slight overprediction of the magnitude of skewness and flatness
in the outer part of the buffer layer.
Some resolution effect is observed on the distribution of the turbulence
kinetic energy production rate and total dissipation rate
($\epsilon_t = \nu \overline{u_i \nabla^2 u_i}$, namely the sum of the viscous dissipation and diffusion terms), 
shown in panel (d).
Not surprisingly, dissipation is most affected being
representative of the small-scale dynamics, and we find that coarser mesh 
resolution in the buffer layer, i.e. lower $j_b$, yields reduction of peak dissipation.
Specifically, assuming the case $j_b=128$ as a reference, we find that the underprediction 
is of about $2.5\%$ for $j_b=4$, $1.0\%$ for $j_b=16$, and $2.0\%$ with the cosine stretching function.
Much smaller effect is found on the production term, which is underestimated by at most 
$0.4\%$ at $j_b=4$, and by less than $1\%$ when using the cosine stretching function.
 
%In summary, the evidence is that changing $j_b$ in the range here considered 
%has negligible effect of the flow statistics for any practical purpose.
%This finding would suggest use of a value of $j_b$ as large as possible, as the mesh
%spacing is smaller throughout the buffer layer.

Computational efficiency in the numerical simulation of wall-bounded flows 
is critically affected by the admissible time step.
Given severe bounds placed by the viscous time step restriction, most DNS 
codes rely on implicit treatment of the viscous terms in all coordinate
directions, or at least in the wall-normal direction~\citep{orlandi_00}.
Treatment of the convective terms is instead typically explicit, and 
computations are time advanced at $O(1)$ CFL number. A notable exception is 
the case of pipe flow, in which the metric singularity yields unnecessarily small time 
step towards the pipe axis, and implicit treatment of the convective terms,
or progressive reduction of the Fourier modes towards the axis, becomes necessary~\citep{boersma_11}. 
In any case, also given that the 
axial time step restriction can be alleviated using a moving reference frame~\citep{bernardini_13},
the wall-normal time step restriction is typically the most restrictive,
and it can be mitigated through suitable design of the stretching function.
\begin{figure}
 \begin{center}
  (a) \includegraphics[width=7.0cm,clip]{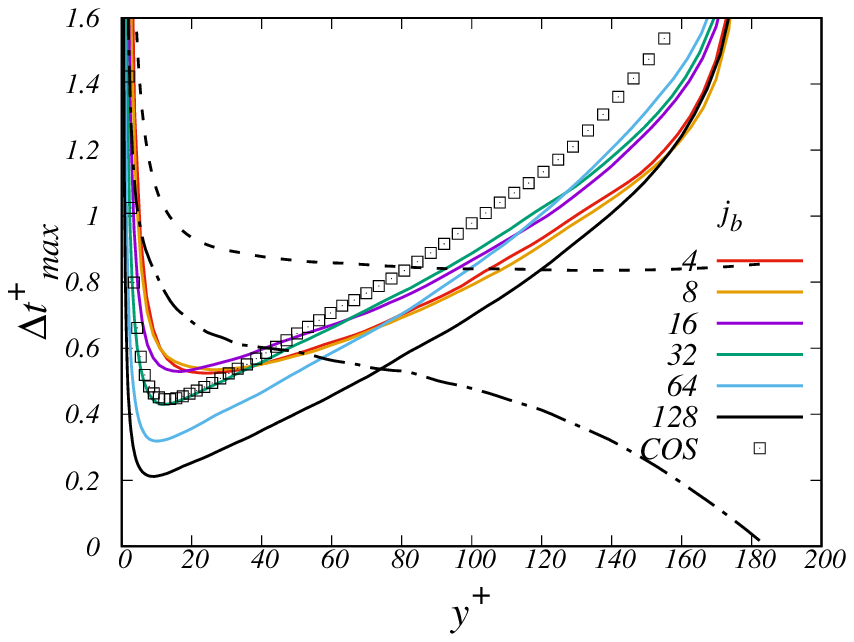}
  (b) \includegraphics[width=7.0cm,clip]{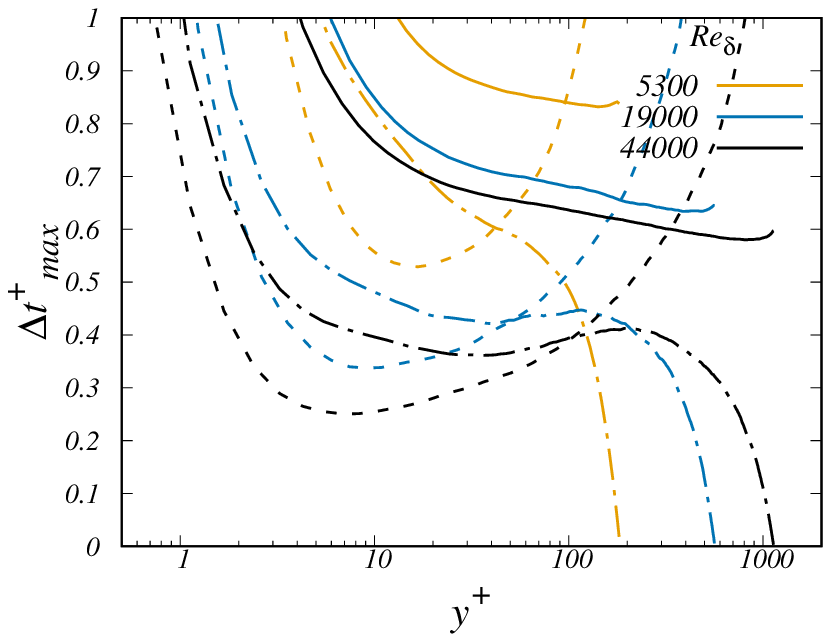}
\caption{Time step limitations in DNS of pipe flow at $\Rey_{\delta}=5300$ as a function of 
$j_b$ (a), and for $j_b=16$ at various $\Rey_{\delta}$ (b). 
The line style indicates the limitation associated with each spatial direction:
wall-normal (solid), axial (dashed), and azimuthal (dash-dotted).
The square symbols denote results obtained using the cosine stretching function, with $N_y=64$.}
  \label{fig:dtmax}
 \end{center}
\end{figure}
This is portrayed in figure~\ref{fig:dtmax}(a), where we show the local convective time step restriction
associated with each coordinate direction, assuming CFL=1,
hence $\Delta t_{\mathrm{max}, z} = \Delta z / (\max | u_z |) $ (axial, solid line),
$\Delta t_{\mathrm{max}, r} = \min ( \Delta y / | u_r |)$ (radial, dashed lines),
$\Delta t_{\mathrm{max}, \theta} = \min ( r \Delta \theta / | u_{\theta} |)$ (azimuthal, dash-dotted line).
The viscous time step limitations are not shown as they can be easily by-passed 
by implicit time integration. Also, since the axial and azimuthal mesh spacings
are not changed, only one dashed line and one dash-dotted lines are shown.
The figure confirms that the axial time step restriction is less compelling
than the other (the simulation is carried out in a moving frame of reference),
and the spanwise restriction becomes too demanding towards the axis, 
but this is disregarded in our simulations as we rely on implicit treatment of the convective terms in
the $\theta$ direction~\citep{akselvoll_96}.
The wall-normal time step restriction (solid lines) is thus found to be the most compelling, 
and to depend critically on the mesh stretching
through the parameter $j_b$, with $\Delta t^+_{\mathrm{max}, r} \approx 0.5$ for $j_b \le 16$,
and reducing to $\Delta t^+_{\mathrm{max}, r} \approx 0.2$, for $j_b=128$. 
 
Given the previously noted improved resolution of the small
scales at increasing $j_b$, it seems that a good compromise between
accuracy and computational efficiency (i.e. large time step) is achieved
for $j_b=16$.  
This value is thus retained in additional tests at higher Reynolds, 
whose results are shown in panel (b). There, results of pipe flow DNS are reported 
for $\Rey_{\delta}=19000$ on a $896 \times 150 \times 896$ mesh (in $z$, $r$, $\theta$, respectively),
and at $\Rey_{\delta}=44000$ on a $1792 \times 270 \times 1792$ mesh.
The same qualitative behavior of the
time step restriction is found at all $\Rey_{\delta}$,
however with reduction of the inner-scaled maximum time step as a consequence of the increased intensity
of vertical velocity fluctuations in the buffer layer.

\subsection{Effect of the $\Delta y^+_w$ parameter}
\label{sec:dyw}

\begin{figure}
 \begin{center}
  (a) \includegraphics[width=7.0cm]{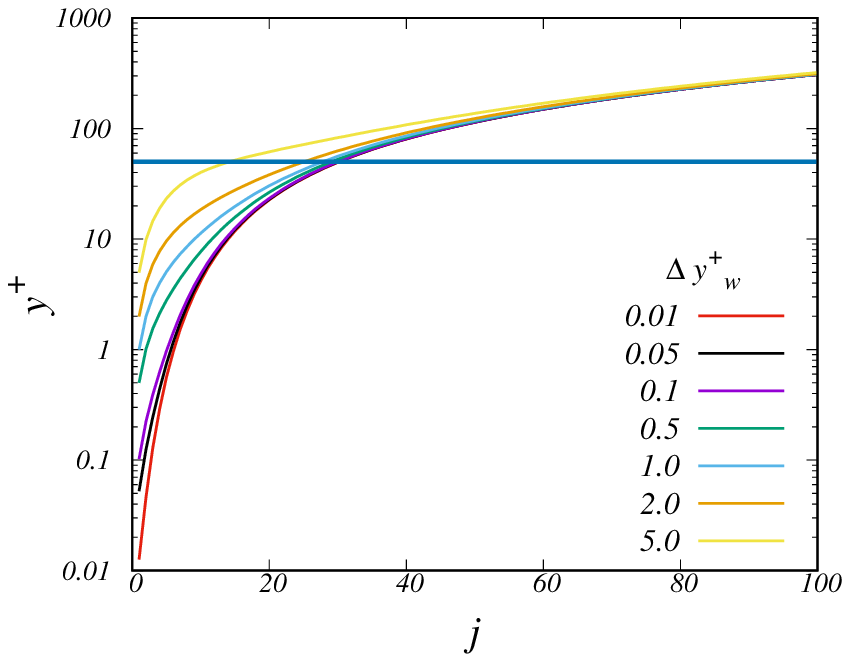}
  (b) \includegraphics[width=7.0cm]{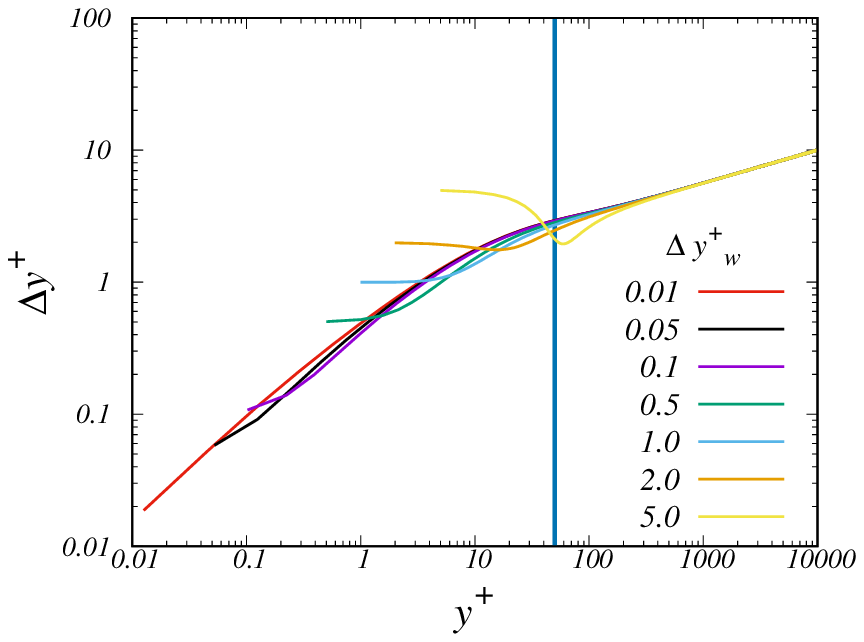}
% (a) \includegraphics[width=7.0cm     ]{FIGURES/yp_new_all.eps}
% (b) \includegraphics[width=7.0cm,clip]{FIGURES/dyp_new_all.eps}
\caption{Mapping functions (a) and corresponding grid spacing distributions (b), 
according to equations~\eqref{eq:mapping}, \eqref{eq:dyp}, as a function of $\Delta y^+_w$,
assuming $j_b=16$, $\alpha=1.25$. 
Straight horizontal and vertical lines denote the edge of the buffer layer ($y^+=50$).} 
  \label{fig:mappings_dyw}
 \end{center}
\end{figure}

The effect of the wall spacing parameter $\Delta y^+_w$, has then been evaluated by 
assuming $j_b=16$, $\alpha=1.25$. 
The resulting stretching functions and the associated grid spacing distributions
are shown for various $\Delta y^+_w$ in figure~\ref{fig:mappings_dyw}.
Whereas the distribution of the grid points outside the buffer layer is essentially the same for all $\Delta y^+_w$,
small values of the wall spacing parameter yield smaller spacing near the wall, and 
larger spacing within the buffer layer. At extreme values ($\Delta y^+_w > 1$),
the grid spacing distribution may even exhibit a reversed trend with respect to the wall distance.

\begin{figure}
 \begin{center}
  (a) \includegraphics[width=7.0cm,clip]{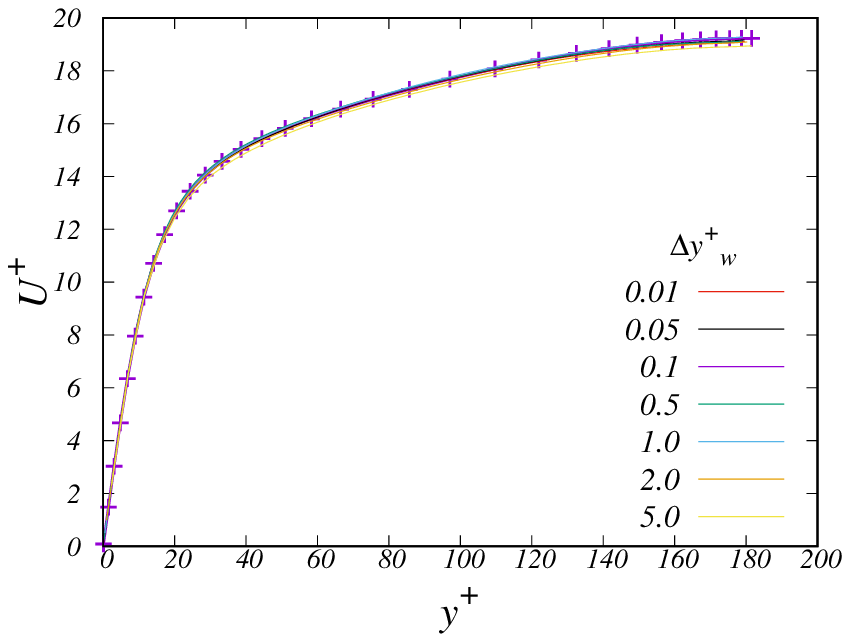}
  (b) \includegraphics[width=7.0cm,clip]{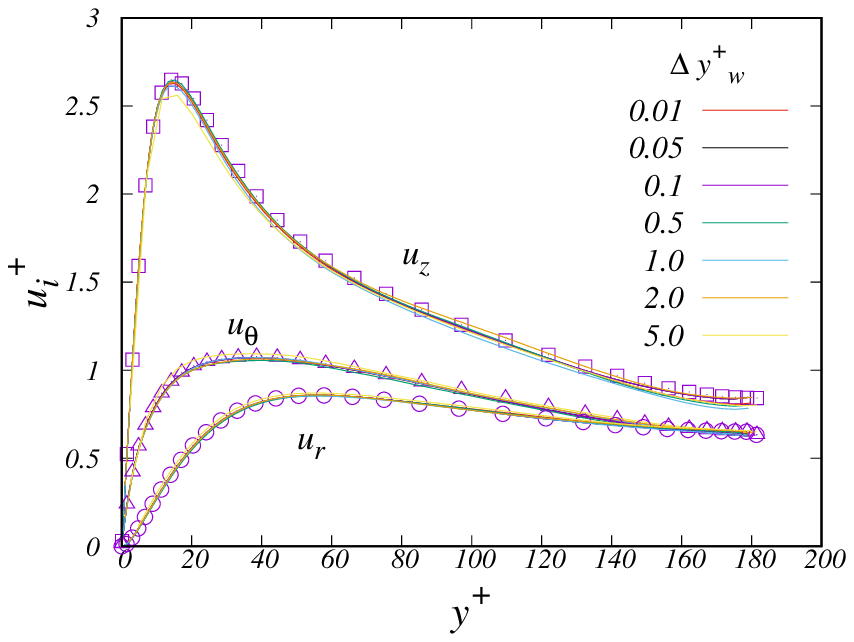} \\
  (c) \includegraphics[width=7.0cm,clip]{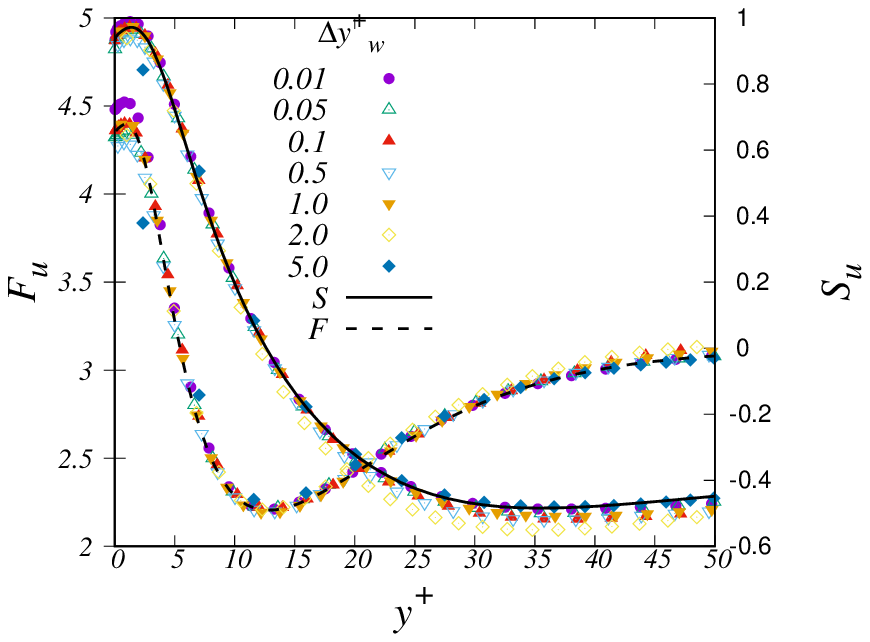}
  (d) \includegraphics[width=7.0cm,clip]{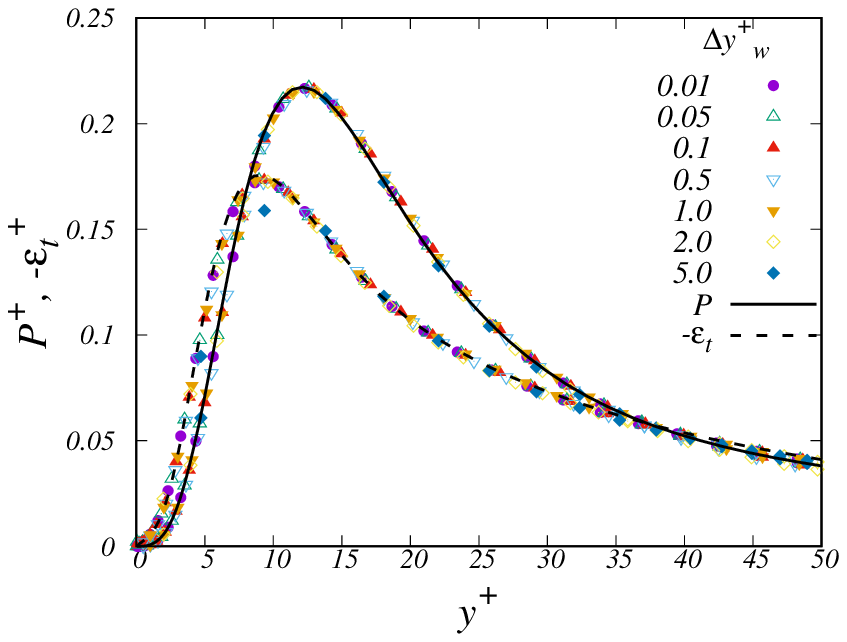}
\caption{DNS of pipe flow at $\Rey_{\delta}=5300$, effect of the $\Delta y^+_w$ wall spacing parameter: 
profiles of (a) mean velocity, (b) r.m.s. velocity fluctuations (axial, $u_z$; wall-normal, $u_r$; azimuthal, $u_{\theta}$), 
(c) skewness and flatness of $u_z$,
and (d) turbulence kinetic energy production ($P=-\overline{u_r u_z} \, \diff U / \diff y$, solid lines),
and total dissipation ($\varepsilon_t = \nu \overline{u_i \nabla^2 u_i}$, dashed lines). 
Symbols in panels (a, b) denote velocity statistics of \citet{wu_08}. 
Lines in panels (c, d) denote numerical results obtained with $j_b=128$, $\Delta y^+_w$, which
are used as a reference. The square symbols denote results obtained using the cosine stretching function, with $N_y=64$.}
  \label{fig:stats_Re5300_dyw}
 \end{center}
\end{figure}

A series of pipe flow simulations have been carried out at $\Rey_b=5300$,
by retaining the same number of grid points in the radial direction, $N_r=67$.
As $\Delta y^+_w$ varies from $0.01$ to $1$, $\Rey_{\tau}$ ranges between $180.9$ and $181.5$,
hence with scatter of less than $0.3 \%$. Deviations become about $1\%$ for $\Delta y^+_w=5$,
with $\Rey_{\tau}=183.2$. Detailed results of the grid sensitivity study are shown in figure~\ref{fig:stats_Re5300_dyw}.
The error in the mean velocity profiles (panel (a)) is limited to well less than $1\%$ for $\Delta y_w^+ \le 2$,
and it is only apparent for $\Delta y_w^+ =5$. The velocity variances (panel (b)) are a bit more
sensitive, and some scatter in the outer layer is apparent already at $\Delta y_w^+ =1$.
The higher-order moments (panel (c)) are most affected by the near-wall resolution, and 
especially the flatness keeps increasing in the viscous sublayer as $\Delta y_w^+$ is reduced. 
Far from the wall, skewness and flatness are very weakly affected, as long as $\Delta y_w^+ \le 1$.
Notably, the total dissipation (panel (d)) is also weakly affected by $\Delta y_w^+$,
with the exception of the case $\Delta y^+_w=5$, which yields a reduced peak value.
The maximum allowed radial time step (not shown) is barely affected for small values of 
the wall spacing parameter, although we find some limited gain with use of $\Delta y^+_w=0.05$, 
and it decreases for $\Delta y^+_w > 0.5$ as a result of smaller grid spacing in the buffer layer.
 
\subsection{Assessment}
\label{sec:test}

\begin{figure}
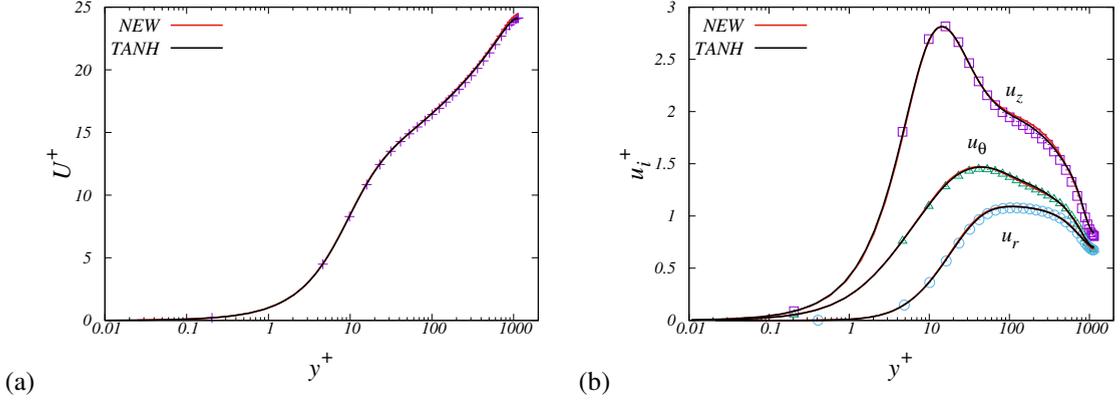

 \begin{center}
  (a) \includegraphics[width=7.0cm,clip]{uplus.eps}
  (b) \includegraphics[width=7.0cm,clip]{urms.eps}
\caption{DNS of pipe flow at $\Rey_{\delta}=44000$: profiles of (a) mean velocity and (b) r.m.s. velocity fluctuations
(axial, $u_z$; wall-normal, $u_r$; azimuthal, $u_{\theta}$), obtained using the new stretching function (with $j_b=16$, $N_y=270$) and a traditional hyperbolic tangent stretching function (with $\beta=3.8$, $N_y=512$). Symbols denote data of \citet{wu_08}.}
  \label{fig:stats_Re44000}
 \end{center}
\end{figure}

As a final assessment of the proposed stretching function, in figure~\ref{fig:stats_Re44000} we show
the velocity statistics obtained at $\Rey_{\delta}=44000$ (yielding $\Rey_{\tau} \approx 1140$)
using the proposed stretching function with $j_b=16$, $\Delta y^+_w=0.05$, $\alpha=1.25$, $N_y=270$. 
As a basis of comparison we consider DNS results obtained with the same code, using a classical 
hyperbolic tangent stretching function~\citep{orlandi_00},
with stretching parameter $\beta=3.8$, and $N_y=512$, which can be regarded as a well resolved DNS.
We also compare with the reference DNS results of \citet{wu_08}. Again, no significant difference arises in the primary
flow statistics, despite the vastly different distribution of grid points. On the other hand, the
time step is $\Delta t^+=0.22$ when using the new stretching, as compared to $\Delta t^+=0.12$ when using hyperbolic tangent stretching, 
with clear reduction of computer time.
Additional assessment of our stretching function is reported in a separate publication,
in which we carry out DNS of pipe flow up to $\Rey_{\tau} \approx 6000$~\citep{pirozzoli_21}.

\section{Conclusions}
\label{sec:conclusions}

It is a fact that, although DNS of wall-bounded flows is by now a well established subject, the choice 
of the wall-normal clustering of grid points is frequently made based on subjective judgement, or 
based on constraints from the numerical algorithm. 
With the purpose of systematizing the matter, we propose a simple stretching function
as given in equation~\eqref{eq:mapping}.
By construction, this mapping has the natural property of yielding precisely constant resolution in terms of the 
local Kolmogorov length scale in the outer part of the wall layer, where turbulence is not far from isotropic.
Consistent with previous DNS, we set the resolution parameter in such a way that the grid spacing is 
$\Delta y = 1.25 \eta$, although it can probably be taken a bit higher and reduce the 
total number of grid points. Interestingly, imposing constant resolution in Kolmogorov units
implies that the number of grid points in the wall-normal direction should scale 
as $\Rey_{\tau}^{3/4}$, hence a bit milder rate than in the wall-parallel directions.
The outer-layer stretching is combined with a near-wall uniformly-spaced distribution by means 
of a blending function which is controlled by a parameter ($j_b$) which may be interpreted as the grid
node index at which transition takes place, thus
larger values of $j_b$ imply a larger number of grid points within the buffer layer.
Another relevant parameter is the wall grid spacing, $\Delta y^+_w$,
controlling resolution in the viscous sublayer.
We have found that the computed buffer-layer statistics at low 
Reynolds number are to a large extent independent of both $j_b$ and $\Delta y^+_w$,
with a sensitivity of $O(1\%)$ at most for properties associated with
small-scale turbulence activity as viscous dissipation.

On the other hand, the allowed computational time step is crucially affected by $j_b$ (much less 
by the wall grid spacing), and compromise between accuracy and efficiency leads us to suggest $j_b=16$, $\Delta y^+_w=0.05$, 
as an optimal set of stretching parameters. 
Numerical simulations carried out for pipe flow at moderate Reynolds number ($\Rey_{\tau} \approx 10^3$) 
support effectiveness and accuracy of the proposed stretching function at higher 
Reynolds number than considered in the preliminary tests. Results obtained
at more extreme Reynolds number are reported elsewhere~\citep{pirozzoli_21}.
Although DNS are shown here only for the case of pipe flow, we believe that the same mapping
(perhaps with slight modifications) can be adapted to study channel flow and boundary layers,
on account of the near-universality of wall-bounded turbulence~\citep{monty_09}. 
We thus trust that the proposed stretching can be profitably used as a common basis for the design
of future DNS to reach and exceed the current threshold of $\Rey_{\tau} \approx 10^4$.

%\section*{Acknowledgments}

\bibliographystyle{model1-num-names}
\bibliography{references}

\end{document}